\documentclass{amsart}
\usepackage{amsmath,amsthm,url,graphicx, latexsym, curves}
\usepackage{amsfonts}
\usepackage{amssymb}

\numberwithin{equation}{section}

\begin{document}

\title[smallest eigenvalue for fixed trace Laguerre beta-ensemble]{
Smallest eigenvalue distribution of the fixed trace Laguerre beta-ensemble}
\author{Yang Chen,
Dang-Zheng Liu \and  Da-Sheng Zhou}
\address{Department of Mathematics, Imperial College London, 180 Queen¡¯s
Gates, London SW7 2BZ, UK}\email{ychen@ic.ac.uk}

\address{School of Mathematical Sciences, Peking University, Beijing,
100871, P.R. China} \email{dzliumath@gmail.com}


\address{Department of Mathematics, University of Macau, Av. Padre Tom\'{a}s
Pereira, Taipa, Macau, P.R. China} \email{zhdasheng@gmail.com}

\begin{abstract}

In this paper we study entanglement of the reduced
density matrix of a bipartite quantum system in a random pure state.

It transpires that this involves the computation of the smallest
eigenvalue distribution of the fixed trace Laguerre ensemble of
$N\times N$ random matrices. We showed that for finite $N$ the
smallest eigenvalue distribution may be expressed in terms of Jack
polynomials.

Furthermore, based on the exact results, we found, a limiting distribution,
when the smallest eigenvalue is suitably scaled with $N$ followed by a large $N$ limit.
Our results turn out to be the same as the smallest eigenvalue distribution of the classical Laguerre ensembles
without the fixed trace constraint. This suggests in a broad sense, the
global constraint does not influence local correlations, at least, in the large $N$ limit.

Consequently, we have solved an open problem:  The determination of the smallest
eigenvalue distribution of the reduced density matrix---obtained by tracing out
the environmental degrees of freedom---for a bipartite quantum system of unequal dimensions.

\end{abstract}

\maketitle




\newcommand{\newc}{\newcommand}
\newc{\beq}{\begin{equation}}
\newc{\eeq}{\end{equation}}
\newc{\kt}{\rangle}
\newc{\br}{\langle}
\newc{\beqa}{\begin{eqnarray}}
\newc{\eeqa}{\end{eqnarray}}
\newc{\pr}{\prime}
\newc{\longra}{\longrightarrow}
\newc{\ot}{\otimes}
\newc{\rarrow}{\rightarrow}
\newc{\h}{\hat}
\newc{\bom}{\boldmath}
\newc{\btd}{\bigtriangledown}
\newc{\al}{\alpha}
\newc{\be}{\beta}
\newc{\ld}{\lambda}
\newc{\ldmin}{\lambda_{\rm min}}
\newc{\sg}{\sigma}
\newc{\p}{\psi}
\newc{\eps}{\epsilon}
\newc{\om}{\omega}
\newc{\mb}{\mbox}
\newc{\tm}{\times}
\newc{\hu}{\hat{u}}
\newc{\hv}{\hat{v}}
\newc{\hk}{\hat{K}}
\newc{\ra}{\rightarrow}
\newc{\non}{\nonumber}
\newc{\ul}{\underline}
\newc{\hs}{\hspace}
\newc{\longla}{\longleftarrow}
\newc{\ts}{\textstyle}
\newc{\f}{\frac}
\newc{\df}{\dfrac}
\newc{\ovl}{\overline}
\newc{\bc}{\begin{center}}
\newc{\ec}{\end{center}}
\newc{\dg}{\dagger}
\newc{\prh}{\mbox{PR}_H}
\newc{\prq}{\mbox{PR}_q}
\newpage
\section{Introduction}
A bipartite quantum system, is a composite system which can be
described as a product state of two subsystems; one of which we
would like to study and the other describing the environment such as
noise or heat baths.

Due to its central role in quantum information
and quantum computation, which is considered as an indispensable
resource, entanglement of a bipartite quantum
system has recently generated a flurry of activities and
has been studied extensively \cite{Neilsen}.\label{sect1}

There are extensive literature on this topic; here we list those
which are of immediate relevance to our paper, see for examples
\cite{Page}, \cite{hall}, \cite{So}, \cite{HLW}, \cite{Nechita},
\cite{mbl} and \cite{lz}. Statistical properties of such random
states are also important in the characterization of quantum chaotic systems, see
\cite{mbl,HLW,lz} and references therein.

A bipartite quantum system consists of a system ($A$) and its
environment ($B$). Explicitly, we consider a composite
system which is described by a $(NM)$-dimensional Hilbert space
$\mathcal{H}^{(NM)}=\mathcal{H}^{(N)}_{A} \otimes
\mathcal{H}^{(M)}_{B}$. Let ${|e_{i}^{A}\rangle }_{i=1}^{N}$ and
${|e_{j}^{B}\rangle }_{j=1}^{M}$ be the complete orthogonal basis
for the subsystems $A$ and $B$, respectively.  Without loss of
generality, we assume that $N\leq M$. Any quantum state
$|\Phi\rangle$ in the Hilbert space $\mathcal{H}^{(NM)}$ can be
expressed as the linear combination of $|e_{i}^{A}\rangle\otimes |e_{j}^{B}\rangle$
as follows:
\begin{equation}
|\Phi \rangle =\sum_{i=1}^{N}\sum_{j=1}^{M}X_{ij}
   \label{linear combibnation},
\end{equation}
where the coefficients $X_{ij}\in \mathbb{C}$ form a rectangular
$N\times M$ complex matrix $X=[X_{ij}]$.

By a random state we understand that $X_{ij}$
are random variables, so that the resulting density matrix of the subsystem $A,$ say, is a random matrix. See
(1.2) below for a definition.

The composite state $|\Phi\rangle $ is fully unentangled or separable
if $|\Phi \rangle $ can be expressed as a direct product of two states
$|\Phi ^{A}\rangle$ and $|\Phi ^{B}\rangle$ drawn from the Hilbert space of
$A$ and $B$ respectively; that is,
$$
|\Phi \rangle =|\Phi ^{A}\rangle
\otimes \Phi ^{B}\rangle,
$$
 otherwise, it is referred to as an
entangled state.

We say that $|\Phi \rangle $ is a normalized pure
state if and only if associated density matrix defined by
$$
\rho =|\Phi \rangle \langle
\Phi |
$$
 satisfies
 $$
 \text{tr\thinspace }[\rho ]=1.
 $$
The reduced density matrix of the subsystem $A$ obtained by tracing
out the states of subsystem $B$ is found to be, after an easy
computation \cite{bengtsson,mbl},
\begin{equation}
\rho _{A}:=\text{tr\thinspace }_{B}[\rho
]=\sum_{j=1}^{M}\langle e_{j}^{B}|\rho |e_{j}^{B}\rangle
=\sum_{i,j=1}^{N}W_{ij}|e_{i}^{A}\rangle \langle e_{j}^{A}|,
\label{reduces matrix}
\end{equation}
where $W_{ij}$ are the entries of $N\times N$ square matrix
$$
W:=XX^{\dag }
$$
with $\text{tr\thinspace }W=1$ implied by the normalization condition that
$\text{tr\thinspace }[\rho ]=1$. The reduce density matrix of subsystem $B$ is similarly defined,
$\rho _{B}=\text{tr\thinspace }_{A}[\rho ].$

The fact that $\text{tr\thinspace }W=1$ implies the fixed trace ensemble to be introduced
below.

Because $XX^{\dag }$ and $X^{\dag }X$ have the same
non-negative eigenvalues, it is not difficult to see that the
reduced density matrices
$\rho _{A}$ and $\rho _{B}$ have the same set of non-negative eigenvalues
$$
\{\lambda_{i}\}_{i=1}^{N}
$$
and  satisfy the fixed trace condition
$$
\sum_{i=1}^{N}\lambda_{i}=1.
$$
Let $v_{i}^{A} $ be the
eigenvector of the square matrix $W$ corresponding to the eigenvalue
$\lambda_{i}$. Then the density matrix of the subsystem $A$
can be expressed as
$$
\rho_{A}=\sum_{i=1}^{N}\lambda_{i}|v_{i}^{A}\rangle \langle v_{i}^{A}|.
$$
A similar representation holds for $\rho _{B}$.
The composite state
$|\Phi \rangle $ has the well-known Schmidt spectral decomposition
\cite{bengtsson}
\begin{equation}
|\Phi \rangle =\sum_{i=1}^{N}\sqrt{\lambda_{i}}|v_{i}^{A}\rangle
\otimes |v_{i}^{B}\rangle .
\end{equation}

As we have discussed above, a pure state is random if the coefficients $X_{ij}$ are random
variables. The simplest and most common choice is to take
$X_{ij}$ to be independent and identically distributed Gaussian random
variables, see \cite{mbl} for detailed exposition. Therefore, the
eigenvalues of the reduced density matrix $\rho _{A}$ are
distributed according to the joint probability density function
(j.p.d.f) of the $N\times N$ complex Wishart matrix whose
trace is constrained to unity, this corresponds to the case where
the ``symmetry'' parameter $\beta$ is 2. By the method
of random matrix theory \cite{mehta}, it is easily shown that the
j.p.d.f of eigenvalues
is given by, for $\ld_j\geq 0,\;\;j=1,2,..,N,$
\beq
P_{\delta}(\ld_1,\ld_2,\cdots,\ld_N)=C_{N,M}
\delta\,\left(\sum_{l=1}^N \ld_l -1 \right) \prod_{i=1}^{N}
\ld_i^{\f{\beta}{2}\alpha} \prod_{j<k} |\ld_j-\ld_k|^\beta,
\label{jpdf1}
\eeq
where
$\alpha=M-N+1-\frac{2}{\beta}>-\frac{2}{\beta}, \beta>0$. In the situations where
$\beta=1,\ 2,\ 4$ we have the real,
complex and quaternion fixed trace Wishart random matrix ensembles.

For general $\beta$, the j.p.d.f. in (\ref{jpdf1})
can be realized  by a tri-diagonal real symmetric matrices (with random entries) of the
form (see \cite{de})
$$
L_\beta=B_\beta
B_\beta^{T}
$$
subject to $\textrm{tr}[L_\beta]=1$, where $B_{\beta}$ is a bi-diagonal
random matrix of the form
\begin{equation}
\label{matrixmodel} B_\beta \sim \begin{pmatrix}
   {\chi }_{\beta M} &  &  & &   \\
{\chi }_{(N-1)\beta} & {\chi }_{\beta M-\beta}& & &  &    \\
  & {\chi }_{(N-2
  )\beta} & {\chi }_{\beta M-2\beta} &   &   \\
 &\;  &\;\ddots & \;\ddots  & &  \\
     &  && {\chi }_{\beta} &{\chi }_{\beta M-(N-1)\beta}
\end{pmatrix}\end{equation}
and the random variable $\chi_{a}$ has the density
$$
\frac{2^{1-a/2}}{\Gamma{(a/2)}}x^{a-1}e^{-x^{2}/2},\ x>0.
$$
The normalization constant, which can be computed in closed form, is given by,
\beq C_{N,M}=\frac{\Gamma(\frac{\beta}{2}MN)
\left[\Gamma\left(1+\frac{\beta}{2}\right)\right]^{N}}
{\prod_{j=0}^{N-1}\Gamma\left(\frac{\beta}{2}(M-j)\right)\Gamma\left(1+\frac{\beta}{2}(N-j)\right)}.
\eeq
See \cite{zs} for a derivation of $C_{N,M}.$

The j.p.d.f. given in (\ref{jpdf1}), for $\beta=2$ can be traced to the work of
Lloyd and Pagels \cite{lloyd} and of Page \cite{Page}.

When $M=N$, namely, the two subsystems Hilbert spaces are of equal dimension, and $\beta=2$, the j.p.d.f.
is referred to in \cite{So} as the
ensemble of random density matrices with respect to the
Hilbert-Schmidt metric in the set
 of all density matrices of size $N$.
The cases where $\beta=1$ and $\beta=4$ are also important because
these describe systems with time-reversal invariance and rotational
symmetry respectively \cite{mehta}.

We assume $\alpha>-\frac{2}{\beta}$, so that $M-N>-1$. This is
chosen so that the normalization constant (1.6) exists.

The study of the eigenvalues of the reduced density matrix
$\rho_{A}=W$ is crucial for the understanding and utilization of
entanglement. In principle, all the information about the spectral
properties of the subsystem $A$, including its degree of
entanglement, is encoded in the j.p.d.f. given by (\ref{jpdf1}). For
example, one classical measure of entanglement, is the von Neumann
entropy defined by
$$
S=-\text{tr\thinspace }\rho_{A} \ln \rho_{A}
=-\sum_{i=1}^{N}\lambda_{i}\ln \lambda_{i}.
$$
The average entropy $\langle S\rangle$---with respect to the j.p.d.f. of (1.4)---
was computed by Page\cite{Page} for large $N$ with $M\geq N$. It was found that,
$$
\langle S\rangle\approx \ln N-\frac{N}{2M},
$$
which shows that a pure subsystem due to its coupling to the environment is more or less  completely
``randomized''.

Another important measure of entanglement is the smallest eigenvalue
$$
\ldmin ={\rm min}(\lambda_1,\lambda_2,\ldots,\lambda_N),
$$
and its distribution function.

This provides, in addition to understanding the nature of entanglement, important
information about the degree with which the effective dimension of
the Hilbert space of the subsystem $A$ can be reduced \cite{mbl}. Indeed,
the average value $\br \ldmin\kt$ of the smallest eigenvalue
was studied  by Znidaric~\cite{znidaric} for the case $N=M$. Based
on the exact $\br \ldmin\kt$ for a small values of $N$, Znidaric
conjectured that $\br \ldmin\kt=1/N^3$ for all $N$ in the complex
case ($\beta=2$). This conjecture was proved in \cite{mbl}  for
$N=M$, both for the complex $(\beta=2)$ and the real ($\beta=1$)
cases. The problem of computing the distribution of $\ldmin$ for
unequal dimensions ($M>N$) was posed in \cite{mbl}, which remains
open and it is this problem that we address.

In this paper we compute the distribution of the smallest eigenvalue
$\lambda_{\rm{min}}$ from the joint probability density function of
(\ref{jpdf1}) for general $\beta$. In particular, for the most
interesting cases where $\beta=1,2,4,$ we will calculate the
distribution of the smallest eigenvalue at finite $N$ and in a large $N$
limit to be described later. In the situation where $\beta=1,$ we assume $M-N$ is odd.

Let us fix the notations to be used throughout the paper.

Let $Q_{N,M}(x)$ be the probability
that $\lambda_{\rm{min}}$ is $\geq$ $x$, that is,
 \beq
 Q_{N,M}(x):=\mbox{Prob}\left[ \ldmin \ge
x \right]= \mbox{Prob}\left[ \ld_{1} \ge x, \ld_2 \ge x, \ldots,
\ld_N \ge x \right], \eeq and the probability density function of
the smallest eigenvalues is of course \beq P_{N,M}(x)=
-\frac{d}{dx}Q_{N,M}(x). \eeq In the situations where $\beta=2,
\alpha=0$ ($M=N$), and $\beta=1, \alpha=-\frac{1}{2}$ ($M=N$),
closed form expressions of $P_{N,M}(x)$ and the moments of $\ldmin$
were obtained in \cite{mbl}. We will extend the above results and
solve a problem posed in \cite{mbl}, namely, the determination of
the probability distribution of $\ldmin$ in the case of unequal
dimensions where $M>N$.


Our results are summarized as follows.

In this paper we extend the known results mentioned above to  the
cases where
\beq \f{\beta}{2}\alpha=m\geq 0,\:\: m\in \mathbb{N} \quad
\:\textrm{and}\;\;\;\beta>0.
 \eeq
We will see that $Q_{N,M}(x)$ and
$P_{N,M}(x)$ are expressed in terms of the Jack symmetric polynomials.

Furthermore, after a scaling
$$
x=\frac{y}{4N^3},
$$
followed by the limit
$N\rightarrow \infty,$ we show that, for a fixed $m,$
\beq
\label{limitdistribution}
Q(y):=
\lim_{N\rightarrow\infty}Q_{N,M}\left(\frac{y}{4N^{3}}\right)
=\exp\left(-\beta \frac{y}{8}\right)\:_{0}F^{(\beta/2)}_{1}
\left(2m/\beta;y_{1},\ldots,y_{m}\right)|_{y_{1}=y_2=...=y_{m}=\frac{y}{4}}
\eeq and
\beq
\label{limitprobabilitydistribution}
P(y):=
\lim_{N\rightarrow\infty}
\frac{1}{4N^{3}}P_{N,M}\left(\frac{y}{4N^{3}}\right)
=A_{m,\beta}\,y^{m}e^{-\beta \frac{y}{8}}\, _{0}F^{(\beta/2)}_{1}
(2m/\beta+2;y_{1},\ldots,y_{m})|_{y_{1}=y_{2}=...=y_m=\frac{y}{4}}
\eeq where \beq A_{m,\beta}=4^{m}(\beta/2)^{\beta/2+2m+1}
\frac{\Gamma(1+\beta/2)}{\Gamma(1+m)\Gamma(1+m+\beta/2)}.
\eeq
Here
$$
_0F_1\left(\frac{2m}{\beta};y_1,...,y_m\right)
$$
is a generalized hypergeometric function.

Therefore the limiting distributions given by
(\ref{limitdistribution}) and (\ref{limitprobabilitydistribution})
are the same as the corresponding results for the Laguerre
beta-ensemble, obtained in \cite{forrester,forrester2} and
\cite{demitriu}. See also the relevant references in \cite{demitriu}.

In the next section, we introduce a multiple integral from the j.p.d.f.
 which ultimately determines the smallest eigenvalue
distribution for $\beta>0.$ In section \ref{beta=2case}, by using an
alternative method of Mehta \cite{mehta} we compute the smallest
eigenvalue distribution for $\beta=2,$ and $\alpha=M-N$ a
non-negative integer, which we later specialize to $\al=2.$ The
conclusion can be found in section \ref{conclusion}.


\section{Distribution of the smallest eigenvalue.}

For convenience let us initially take the trace to be $t,$ where
$t>0.$ We define a function of $x$ and $t$ as follows:
 \beq
 I(x,t):=
\int_{[x,\infty)^{N}} \delta\left(\sum_{i=1}^N \ld_i -t
\right)\prod_{i=1}^N \ld_i^{\f{\beta}{2}\alpha}\, \prod_{1\leq
j<k\leq N} |\ld_j-\ld_k|^\beta\, d^{N}\lambda, \label{qmin}
\eeq
where $d^{N}\lambda:=d\lambda_1\cdots d\lambda_N.$
Clearly, the
distribution $Q_{N,M}(x)$ is given by
\beq
Q_{N,M}(x)=\frac{I(x,1)}{I(0,1)}.
\eeq
With the parameter $t$ we may now take a Laplace transform of $I(x,t)$ with respect to $t.$

Clearly, the numerator of (2.2) is obtained by a Laplace transform of $I(x,t)$
with respect to $t$ and followed by setting $t=1$ after a Laplace inversion.
The denominator of (2.2) is  $1/C_{N,M}$, with $C_{N,M}$ as in (1.6).

To proceed, we first take the Laplace transform $I(x,t)$ with respect to $t$ and compare the results to those of  Laguerre
$\beta$-ensemble, which has been well studied. Thus
\beq
\int_0^{\infty} I(x,t) e^{-st} dt = \int_{[x,\infty)^N} e^{-s
\sum_{i=1}^{N} \ld_i} \prod_{i=1}^N \ld_i^{\f{\beta}{2}\alpha}\,
\prod_{1\leq j<k\leq N} |\ld_j-\ld_k|^\beta\, d^{N}\lambda.
\eeq
After a linear shift followed a scale change,
$$
z_i=\frac{2s}{\beta}\Big(\ld_i-x\Big)\;\;{\rm or\;\;\;} \ld_i=x+\frac{\beta}{2s}z_i,
\quad i=1,...,N
$$
we find,
 \beq
 \int_0^{\infty} I(x,t)
e^{-st} dt =
\left(\frac{\beta}{2s}\right)^{\frac{\beta}{2}MN}e^{-sNx}
J\left(\frac{2s}{\beta}x,\f{\beta}{2}\alpha\right)
\eeq
where
\beq
J(x,\gamma):= \int_{[0,\infty)^N}
\prod_{i=1}^N (x+z_i)^{\gamma}\,e^{-\frac{\beta}{2}z_i} \prod_{1\leq j<k\leq N}
|z_j-z_k|^\beta\, d^{N}z.
\eeq
An inverse Laplace transform leaves
\beq
I(x,t)=\mathcal{L}^{-1}\Big[\Big(\frac{\beta}{2s}\Big)^{\frac{\beta}{2}MN}e^{-sNx}
J\Big(\frac{2s}{\beta}x,\f{\beta}{2}\alpha\Big)\Big](t).
 \eeq
Hence,
\begin{align}
Q_{N,M}(x)&=\frac{I(x,1)}{I(0,1)}=\frac{\mathcal{L}^{-1}\big[\big(\frac{\beta}{2s}\big)^{\frac{\beta}{2}MN}
e^{-sNx}J\big(\frac{2s}{\beta}x,\f{\beta}{2}\alpha\big)\big](1)}
{\mathcal{L}^{-1}\Big[\big(\frac{\beta}{2s}\big)^{\frac{\beta}{2}MN}
J\big(0,\f{\beta}{2}\alpha\big)\Big](1)}\nonumber\\
&=\frac{1}{\mathcal{L}^{-1}\Big[s^{-\frac{\beta}{2}MN}\Big](1)}\mathcal{L}^{-1}\Big
[\frac{e^{-sNx}}{s^{\frac{\beta}{2}MN}}\frac{J\big(\frac{2s}{\beta}x,\f{\beta}{2}\alpha
\big)}{J\big(0,\f{\beta}{2}\alpha\big)}\Big](1)\nonumber\\
&=\Gamma\Big(\frac{\beta}{2}MN\Big)\:\:\mathcal{L}^{-1}\Big
[\frac{e^{-sNx}}{s^{\frac{\beta}{2}MN}}\frac{J\big(\frac{2s}{\beta}x,\f{\beta}{2}\alpha
\big)}{J(0,\f{\beta}{2}\alpha )}\Big](1)\label{relationofILT2}\\
&=\Gamma\Big(\frac{\beta}{2}MN\Big)\mathcal{L}^{-1}\Big
[\frac{1}{s^{\frac{\beta}{2}MN}}\frac{J\big(\frac{2s}{\beta}x,\f{\beta}{2}\alpha
\big)}{J(0,\f{\beta}{2}\alpha )}\Big](1-N x), \label{relationofILT2}
\end{align}
where we have made used the following properties of the inverse Laplace
transform:
\begin{flalign}\label{propertyofILT1}
\mathcal{L}^{-1}\big[s^{-a}\big](t)=\frac{t^{a-1}}{\Gamma(a)}\,\theta(t),\quad
\Re(a)>0,
\end{flalign}
and
\begin{flalign}\label{propertyofILT2}
\mathcal{L}^{-1}\big[f(s)e^{\sigma}\big](t)=\mathcal{L}^{-1}\big[f(s)\big](\sigma+t).
\end{flalign}
In (2.9) the Heaviside function $\theta(t)$ is $1$ for $t>0$ and
$\theta(t)$ is $0,$ for $t<0.$

We should like to mention that the numerator of
\beq \label{denominatSelberg integral}
\frac{J\big(x,\f{\beta}{2}\alpha\big)}{J(0,\f{\beta}{2}\alpha\big)}
\eeq
is an important but difficult multiple integral, while the denominator can be evaluated as a particular Selberg integral.

Observe that (\ref{denominatSelberg integral}) can be interpreted as
the moment of order ($\f{\beta}{2}\alpha $) of the characteristic
polynomial of the equivalent tri-diagonal   matrix model of Laguerre
$\beta$-ensembles \cite{de}. In particular, for $\beta=1,2,4,\;$ $J(x,\alpha\beta/2)$ can
be evaluated in closed form in  terms of Laguerre polynomials, see Chapter 22 of \cite{mehta}.


We recall for the reader the definition of the generalized
hypergeometric function with positive parameter $\nu$. This arises in an extensive investigation by Kaneko
\cite{kaneko} on the multi-variable version of the Amoto's generalization of the Selberg integral.
See also \cite{forrester2}, \cite{demitriu}.

The generalized hypergeometric function of $m$ variables is as follows:
\beq _{p}F^{(\nu)}_{q}
(a_{1},\ldots,a_{p} ;b_{1},\ldots,b_{q} ;x_{1},\ldots,x_{m}): =
\sum_{k=0}^{\infty}\sum_{|\kappa|=k}\frac{[a_{1}]^{(\nu)}_{\kappa}\cdots
[a_{p}]^{(\nu)}_{\kappa}}{[b_{1}]^{(\nu)}_{\kappa}\cdots
[b_{q}]^{(\nu)}_{\kappa}}C_{\kappa}^{(\nu)}
(x_{1},\ldots,x_{m}),\label{hypergeometrifunction}
\eeq
where the sum over
$|\kappa|=k$ in (\ref{hypergeometrifunction})  is the sum over
all partitions $(\kappa_{1},\ldots,\kappa_{m})$ of non-negative
integers such that
\beq \kappa_{1}\geq\ldots \geq \kappa_{m},\;\;\;
\sum_{j=1}^{m} \kappa_{j}=k.
\eeq
The generalized  factorial
function $[a]^{(\nu)}_{\kappa}$ is defined  by
\beq
[a]^{(\nu)}_{\kappa}:=\prod_{j=1}^{m}
\left(a-\frac{1}{\nu}(j-1)\right)_{\kappa_{j}},
\eeq
with
\beq
(a)_{k}=a(a+1)\cdots (a+k-1).
\eeq
The  $m-$variable function
$C_{\kappa}^{(\nu)}(x_{1},\ldots,x_{m})$ is a homogeneous symmetric
polynomial  of degree $k$ and it is  proportional  to the so-called
Jack symmetric polynomials, see \cite{kaneko}. A classic reference
for Jack symmetric polynomials is \cite{stanley}.

In the one variable case, where $m=1$, $C_{\kappa}^{(\nu)} (x)=x^{k},$ the generalized hypergeometric
function reduces the one-variable hypergeometric function $\:_{p}F_{q}(x)$.

For non-negative integer $m$ , Forrester \cite{forrester2}
showed that the quotient of (\ref{denominatSelberg integral})
can be expressed as a terminating generalized hypergeometric
function ,
 \beq
\frac{J(x,m )}{J(0,m)}=\, _{1}F^{(\beta/2)}_{1}
(-N;2m/\beta;x_{1},\ldots,x_{m})|_{x_{1}=x_{2}=...x_{m}=-x},
 \eeq
 where $ m=\f{\beta}{2}\alpha$ is a nonnegative integer.
Using the above expression, we obtain from
(\ref{relationofILT2}) that
\begin{align}
&Q_{N,M}(x)\nonumber\\
&=\Gamma\Big(\frac{\beta}{2}MN\Big)\mathcal{L}^{-1}\Big
[\frac{1}{s^{\frac{\beta}{2}MN}}\, _{1}F^{(\beta/2)}_{1}
(-N;2m/\beta;x_{1},\ldots,x_{m})|_{x_{j}=-2sx/\beta}\Big](1-N x)\nonumber\\
&=\Gamma\Big(\frac{\beta}{2}MN\Big)\sum_{k=0}^{\infty}\frac{1}{k!}\sum_{|\kappa|=k}
\frac{[-N]^{(\beta/2)}_{\kappa}}{[2m/\beta]^{(\beta/2)}_{\kappa}}\mathcal{L}^{-1}\Big[C_{\kappa}^{(\beta/2)}
(x_{1},\ldots,x_{m})|_{x_{1}=\cdots=x_{m}=-2x/\beta}
\:\frac{s^{k}}{s^{\beta MN/2}}\Big](1-N x)
\nonumber\\
&=\sum_{k=0}^{\infty}\sum_{|\kappa|=k}\frac{\Gamma\Big(\frac{\beta}{2}MN\Big)}{\Gamma\Big(\frac{\beta}{2}MN-k\Big)}
\frac{[-N]^{(\beta/2)}_{\kappa}}{[2m/\beta]^{(\beta/2)}_{\kappa}}\Big(-\frac{2x}{\beta}\Big)^{k}C_{\kappa}^{(\beta/2)}
(1^{m})\frac{1}{k!}(1-N x)^{\frac{\beta}{2}MN-k-1}\theta(1-Nx)
\nonumber\\
&=\sum_{k=0}^{\infty}\sum_{|\kappa|=k}\Big(-\frac{2}{\beta}\Big)^{k}\frac{\Gamma\Big(\frac{\beta}{2}MN\Big)}
{\Gamma\Big(\frac{\beta}{2}MN-k\Big)}
\frac{[-N]^{(\beta/2)}_{\kappa}}{[2m/\beta]^{(\beta/2)}_{\kappa}}\frac{C_{\kappa}^{(\beta/2)}
(1^{m})}{k!}x^{k}(1-N x)^{\frac{\beta}{2}MN-k-1}\theta(1-Nx)
\label{exactdensity}.
\end{align}
Here $1^{m}=(1,\ldots,1)$ where the number of variables is $m$. We
emphasize that (\ref{exactdensity}) is a finite sum.

Therefore, for finite $N,$ we obtained an expression of the
smallest eigenvalue in terms of Jack polynomials evaluated at $(1,1,...,1)$ that all the $m$
entries are unity.

Next, we compute the moments of $\ldmin$ and its limiting distribution after scaling
with $N.$
\newpage
\textit{Moments of $\ldmin.$}

From the explicit expression of the
distribution in (\ref{exactdensity}), one can easily compute all
the moments of $\ldmin$. For the $p$-th moment, $p\geq 1$,
we have that,
\begin{align}
\mu_p&:= \br \ldmin^p \kt=-\int_0^{\infty} x^p
d\,Q_{N,M}(x)= p\int_0^{\infty} x^{p-1}Q_{N,M}(x) d\,x\nonumber\\
&=p\sum_{k=0}^{\infty}\sum_{|\kappa|=k}\Big(-\frac{2}{\beta}\Big)^{k}
\frac{\Gamma\Big(\frac{\beta}{2}MN\Big)\Gamma(p+k)}{\Gamma\Big(\frac{\beta}{2}MN+p\Big)N^{p+k}}
\frac{[-N]^{(\beta/2)}_{\kappa}}{[2m/\beta]^{(\beta/2)}_{\kappa}}\frac{C_{\kappa}^{(\beta/2)}
(1^{m})}{k!}. \label{moments} \end{align}

\textit{Limit distribution of $\ldmin$}.

We now re-scale $x$ so that $x=\frac{y}{4 N^{3}}.$
Consider a $\kappa$. If  $m$ is fixed, that is, $M-N$ does not depend on $N$, a
simple computation produces two useful formulas:
\begin{align}\label{useful formula1}
  \frac{\Gamma\Big(\frac{\beta}{2}MN\Big)}{\Gamma\Big(\frac{\beta}{2}MN-k\Big)}
\frac{[-N]^{(\beta/2)}_{\kappa}}{N^{3}} \longrightarrow
\Big(-\frac{\beta}{2}\Big)^{k},\quad N\to\infty
\end{align}
and
\begin{align}\label{useful formula2}
  \left(1-\frac{y}{4N^{2}}\right)^{\frac{\beta}{2}MN-k-1}\longrightarrow
  e^{-\beta \frac{y}{8}},\quad N\to\infty.
\end{align}

With (\ref{useful formula1}) and
(\ref{useful formula2}), we find,
\begin{align}
&Q(y)= \lim_{N\rightarrow\infty}Q_{N,M}\Big(\frac{y}{4N^{3}}\Big)\nonumber\\
&=\sum_{k=0}^{\infty}\sum_{|\kappa|=k}
\frac{1}{[2m/\beta]^{(\beta/2)}_{\kappa}}\frac{C_{\kappa}^{(\beta/2)}
(1^{m})}{k!}e^{-\beta
\frac{y}{8}}\Big(\frac{y}{4}\Big)^{k}=e^{-\beta \frac{y}{8}}\,
_{0}F^{(\beta/2)}_{1}
(2m/\beta;y_{1},\ldots,y_{m})|_{y_{1}=\cdots=y_{m}=\frac{y}{4}}
\label{limitdensity}.
\end{align}
 We see that our smallest eigenvalue distribution with the fixed trace constraint,
 namely, (\ref{limitdensity}) is the same as that in \cite{forrester2} for
the Laguerre $\beta$ ensemble without the fixed trace constraint.  Our result seems to
indicate that global constraint does not influence local eigenvalue distribution
at least after suitable scaling followed by a $N\to\infty$ limit.

Performing  a similar computation, we find,
\begin{align}
P(y)&=\lim_{N\rightarrow
\infty}\frac{1}{4N^{3}}P_{N,M}\Big(\frac{y}{4N^{3}}\Big)=-Q'(y)\nonumber\\
&=A_{m,\beta}\,y^{m}e^{-\beta\frac{y}{8}}\, _{0}F^{(\beta/2)}_{1}
(2m/\beta+2;y_{1},\ldots,y_{m})|_{y_{1}=y_{2}=...=y_{m}=\frac{y}{4}},\label{limitsmallestdistr}
\end{align}
where \beq
A_{m,\beta}=4^{m}(\beta/2)^{\beta/2+2m+1}\frac{\Gamma(1+\beta/2)}{\Gamma(1+m)\Gamma(1+m+\beta/2)}.
\eeq
Note that since $P(y)=-Q'(y)$,
(\ref{limitsmallestdistr}) follows from the relation between exact
expressions of $P(y)$ and $Q(y)$ in \cite{forrester2}.

In particular, when $m=0$, we find $Q(y)=e^{-\beta \frac{y}{8}}.$

When $m=1,$ we have,
$$
Q(y)=2^{-1+\frac{2}{\beta}}\Gamma(\frac{2}{\beta})e^{-\beta
\frac{y}{8}}y^{\frac{1}{2}-\frac{1}{\beta}}I_{\frac{\beta}{2}-1}(\sqrt{y}).
$$
Here  $I_{\rho}(x)$ denotes the modified Bessel function of the
first kind \cite{aar}.

\section{another Exact expression at $\beta=2$}
\label{beta=2case}
In this section, we assume that $\beta=2$ and $
\alpha=M-N$ is a nonnegative integer. For $\beta=1$ or $4$, similar
but more sophisticated results in Chapter 22 of \cite{mehta} can be
used to obtain the corresponding results in this section.

Our main result is  stated below:
\begin{align}
Q_{N,M}(x)=\Gamma(MN)x^{MN-1}\mathcal{L}^{-1}\Big
[\frac{1}{s^{MN}}\det\Big[L_{N+k-l}^{(l)}(-s)\Big]_{k,l=0}^{\alpha-1}\Big]
\Big(\frac{1-Nx}{x}\Big).
\end{align}
Here $L_{N}^{(\rho)}(x)$ is a Laguerre polynomial given  by
 \beq L_{N}^{(\rho)}(x)=\sum_{j=0}^{N}\left(\begin{array}{ccc}
N+\rho\\
N-j
\end{array}\right)
\frac{(-x)^{j}}{j!}.
\eeq
See the standard reference \cite{szego}
for Laguerre polynomials.

Again we begin with Eq. (\ref{relationofILT2}), but with $\beta=2.$
We find
\begin{align}\label{distrat2}
Q_{N,M}(x)&=\Gamma(MN)\mathcal{L}^{-1}\Big
[\frac{1}{s^{MN}}\frac{J(s x,\alpha )}{J(0,\alpha )}\Big](1-N
x)\nonumber\\
&=\Gamma(MN)\frac{J(0,0 )}{J(0,\alpha )}\mathcal{L}^{-1}\Big
[\frac{1}{s^{MN}}\frac{J(s x,\alpha )}{J(0,0 )}\Big](1-N x).
\end{align}
The integral (Eq.(17.6.5), Page 321, \cite{mehta}) implies \beq
J(0,\alpha )=\prod_{j=1}^{N}\Gamma(1+j)\Gamma(\alpha+j), \eeq and
hence \beq \label{ratio1}\frac{J(0,0 )}{J(0,\alpha
)}=\prod_{j=1}^{N}\frac{\Gamma(j)}{\Gamma(\alpha+j)}=\left(\prod_{j=1}^{N}(j)_{\alpha}\right)^{-1}.
\eeq On the other hand, it follows from Eq.(22.2.28), Page 416,
\cite{mehta}, that
\beq
(-1)^{\alpha N}\frac{J(-x,\alpha
)}{J(0,0)}=\left(\prod_{l=0}^{\alpha-1}l!\right)^{-1}
\det\Big[\frac{d^{l}}{d x^{l}}C_{N+k}(x)\Big]_{k,l=0}^{\alpha-1}
\eeq
where $C_j(x)$ is the monic polynomial related to the Laguerre
polynomials as follows:
\beq C_{j}(x)=(-1)^{j}j!L_{j}^{(0)}(x).
\eeq

 By an elementary differential-difference relation satisfied by the Laguerre
polynomials,
 \beq \frac{d}{d
x}L_{j}^{(\rho)}(x)=-L_{j-1}^{(\rho+1)}(x),
\eeq
we obtain
\begin{align}
(-1)^{\alpha N}\frac{J(-x,\alpha )}{J(0,0
)}&=\left(\prod_{l=0}^{\alpha-1}l!\right)^{-1}\prod_{l=0}^{\alpha-1}(-1)^{N+l}(N+l)!\det\Big[\frac{d^{l}}{d
x^{l}}L_{N+k}^{(0)}(x)\Big]_{k,l=0}^{\alpha-1}\nonumber\\
&=(-1)^{\alpha
N}\prod_{l=0}^{\alpha-1}\frac{(N+l)!}{l!}\det\Big[L_{N+k-l}^{(l)}(x)\Big]_{k,l=0}^{\alpha-1}\nonumber\\
&=(-1)^{\alpha
N}\prod_{l=0}^{\alpha-1}(l+1)_{N}\det\Big[L_{N+k-l}^{(l)}(x)\Big]_{k,l=0}^{\alpha-1}\label{ratio2}.
\end{align}
Note that
\beq
\label{ratio3}
\prod_{j=1}^{N}(j)_{\alpha}=\prod_{l=0}^{\alpha-1}(l+1)_{N}.
\eeq
Combining Eqs. (\ref{ratio1}), (\ref{ratio2}) and (\ref{ratio3}), we obtain,
\begin{align}
\frac{J(-x,\alpha )}{J(0,\alpha
)}=\det\Big[L_{N+k-l}^{(l)}(x)\Big]_{k,l=0}^{\alpha-1}.
\end{align}
Furthermore, from (\ref{distrat2}), we arrive at the expression,
\begin{align}
Q_{N,M}(x)&=\Gamma(MN)\mathcal{L}^{-1}\Big
[\frac{1}{s^{MN}}\det\Big[L_{N+k-l}^{(l)}(-sx)\Big]_{k,l=0}^{\alpha-1}\Big](1-N
x)\nonumber\\
&=\Gamma(MN)x^{MN}\mathcal{L}^{-1}\Big
[\frac{1}{(sx)^{MN}}\det\Big[L_{N+k-l}^{(l)}(-sx)\Big]_{k,l=0}^{\alpha-1}\Big](1-N
x)\nonumber\\
&=\Gamma(MN)x^{MN-1}\mathcal{L}^{-1}\Big
[\frac{1}{s^{MN}}\det\Big[L_{N+k-l}^{(l)}(-s)\Big]_{k,l=0}^{\alpha-1}\Big]
\Big(\frac{1-Nx}{x}\Big)
\label{exactdistrat2},
\end{align}
where we have used a property  of inverse Laplace transform:
\begin{align}\label{Property1ILT}
\mathcal{L}^{-1}\Big[\frac{1}{b}f\Big(\frac{s}{b}\Big)\Big](t)=\mathcal{L}^{-1}[f(s)](bt).\nonumber
\end{align}

%

Now we focus on the case $ \alpha=M-N=2$. Note that since,
\begin{align}
L_{N}^{(\rho)}(-s)=\frac{(\rho+1)_{N}}{N!}\sum_{j=0}^{N}\frac{(-N)_{j}}{(\rho+1)_{j}}\frac{(-s)^{j}}{j!},
 \end{align}
we have
\begin{align}
&\det\big[L_{N+k-l}^{(l)}(-s)\big]_{k,l=0}^{1}=L_{N}^{(0)}(-s)L_{N}^{(1)}(-s)-L_{N+1}^{(0)}(-s)L_{N-1}^{(1)}(-s)\nonumber\\
&=(N+1)\sum_{i=0}^{N}\frac{(-N)_{i}}{(1)_{i}}\frac{(-s)^{i}}{i!}\sum_{j=0}^{N}\frac{(-N)_{j}}{(2)_{j}}\frac{(-s)^{j}}{j!}
-N\sum_{i=0}^{N+1}\frac{(-N-1)_{i}}{(1)_{i}}\frac{(-s)^{i}}{i!}\sum_{j=0}^{N-1}\frac{(-N+1)_{j}}{(2)_{j}}\frac{(-s)^{j}}{j!}
\nonumber\\
&=\sum_{i}\sum_{j}\frac{(-s)^{i+j}}{i!j!}
\frac{(N+1)(-N)_{i}(-N)_{j}-N(-N-1)_{i}(-N+1)_{j}}{(1)_{i}(2)_{j}}\nonumber\\
&=\sum_{i}\sum_{j}\frac{(-s)^{i+j}}{i!j!}
\frac{(-N)_{i}(-N)_{j}}{(1)_{i}(2)_{j}}\frac{(N+1)(1+j-i)}{N+1-i}.
\end{align}
Here we have used the fact
$$
(N+1)(-N)_{i}(-N)_{j}-N(-N-1)_{i}(-N+1)_{j}=(-N)_{i}(-N)_{j}\frac{(N+1)(1+j-i)}{N+1-i}.
$$
In addition, since we are only interested the case of very large $N,$
we can always write the sum in the form of
(3.14) for fixed $i,j$.  Thus, it follows from
Eq.(\ref{exactdistrat2}) that
\begin{align}
&Q_{N,M}(x)\nonumber\\
&=\Gamma(MN)x^{MN-1} \sum_{i}\sum_{j}\frac{(-1)^{i+j}}{i!j!}
\frac{(-N)_{i}(-N)_{j}}{(1)_{i}(2)_{j}}
\frac{(N+1)(1+j-i)}{N+1-i}\mathcal{L}^{-1}\Big
[\frac{1}{s^{MN-i-j}}\Big]\Big(\frac{1-N x}{x}\Big)\nonumber\\
&=\Gamma(MN)x^{MN-1} \sum_{i}\sum_{j}\frac{(-1)^{i+j}}{i!j!}
\frac{(-N)_{i}(-N)_{j}}{(1)_{i}(2)_{j}}
\frac{(N+1)(1+j-i)}{N+1-i}\frac{\theta(\frac{1-N
x}{x})}{\Gamma(MN-i-j)}
\Big(\frac{1-N x}{x}\Big)^{MN-1-i-j}\nonumber\\
&=(1-N x)^{MN-1} \sum_{i,j}\frac{(-1)^{i+j}}{i!j!}
\frac{(-N)_{i}(-N)_{j}}{(1)_{i}(2)_{j}}
\frac{(N+1)(1+j-i)}{N+1-i}\frac{\Gamma(MN)}{\Gamma(MN-i-j)}
\frac{x^{i+j}}{\Big(1-N x\Big)^{i+j}}\label{explicitsum}.
\end{align}

Setting $x=\frac{y}{4N^{3}}$. Consider a fixed $i,j.$
We now deal with the limit of the individual factors on the right-hand side of
Eq.(\ref{explicitsum}) as $N\longrightarrow \infty$.

It is easily seen that (with $x=\frac{y}{4N^3}$), as $ N\to\infty$,
$$
\left(1-Nx\right)^{MN-1} \longrightarrow e^{-y/4},
$$
$$
\big(1-Nx\big)^{i+j}\longrightarrow  1,
$$
\begin{align}
&\frac{(-N)_{i}(-N)_{j}}{N^{i+j}}\longrightarrow  (-1)^{i+j},\nonumber\\
&\frac{\Gamma(MN)}{\Gamma(MN-i-j)}\frac{1}{N^{2i+2j}}\longrightarrow
1. \nonumber
\end{align}
Hence,
\begin{align} &\lim_{N\rightarrow
\infty}Q_{N,M}(\frac{y}{4N^{3}})\nonumber\\
&=e^{-y/4} \sum_{i,j=0}^{\infty}\frac{(y/4)^{i+j}}{i!j!}
\frac{1+j-i}{(1)_{i}(2)_{j}}\nonumber\\
&=e^{-y/4}
\sum_{i,j=0}^{\infty}\frac{(y/4)^{i+j}}{i!j!}\left(\frac{1}{i!j!}-\frac{1}{(i-1)!(j+1)!}
\right)\nonumber\\
&= e^{-y/4}
\sum_{i=0}^{\infty}\frac{(y/4)^{i}}{i!i!}\sum_{j=0}^{\infty}\frac{(y/4)^{j}}{j!j!}-
 \frac{y}{4}e^{-y/4}
\sum_{i=1}^{\infty}\frac{(y/4)^{i-1}}{(i-1)!i!}\sum_{j=0}^{\infty}\frac{(y/4)^{j}}{j!(j+1)!}
\nonumber\\
&=e^{-y/4}\left(I^{2}_{0}(\sqrt{y})-I^{2}_{1}(\sqrt{y}) \right),
\end{align}
where the modified Bessel function of the first kind is given by
\beq
I_{\rho}(x)=(x/2)^{\rho}\sum_{k=0}^{\infty}\frac{(x/2)^{2k}}{k!\Gamma(\rho+k+1)}.
\eeq

\section{Conclusion.}\label{conclusion}
In this paper we have studied the exact probability distribution of the smallest
eigenvalue of the density matrix of an entangled random pure state. This turns out to be the
same as the smallest eigenvalue distribution of the fixed trace Laguerre random matrix ensemble.
We obtained, for a bipartite quantum system composed of two subsystems
whose Hilbert spaces have dimensions $M$
and $N$ respectively, the exact smallest
eigenvalue distribution for all $M\geq N$ (in the $\beta=1$ case we assume $M- N$ is an odd integer).
The distributions are expressed as the evaluation of a certain hypergeometric functions at special points and
solve an open problem in quantum entanglement.

Our result not only provides important information on  entanglement
of a bipartite quantum system in a random pure state, but also
demonstrates the intimate relations between entanglement of
bipartite quantum systems and the fixed trace Laguerre ensemble. We
may conclude based on our results that in a broad sense the global
constraint does not influence local correlations at least in a
certain large $N$ limit. An indication that this may have wider
validity can be found in \cite{lz} where the authors studied the
kernel and the $n-$ point correlation functions of the fixed and
bounded trace Laguerre ensembles. There it was found that the
suitably scaled kernel in the large $N$ limit converges to the
unrestricted kernel in the bulk, hard and soft edges, using the
terminology of Tracy and Widom. However, we should like to mention
that the ``matching up'' of the kernels in the constrained and the
unrestricted ensembles does not imply that the ``matching up'' of the
corresponding distribution functions (in this instance the smallest eigenvalue distributions).
Only after a rather elaborate computation shows that this is the case.

 Although there is currently no obviously
effective characterization of the degree of entanglement, the von
Neumann entropy, however, is considered to be useful as a
measurement of entanglement in bipartite quantum systems. The
distribution function of the von Neumann entropy may be successfully
tackled using our approach.

\end{document}